\documentclass[lettersize,journal]{IEEEtran}
\usepackage[colorlinks,urlcolor=blue,linkcolor=blue,citecolor=blue]{hyperref}
\usepackage{color,array}
\usepackage{listings}
\usepackage[linesnumbered,ruled,vlined]{algorithm2e}
\usepackage{algorithmic}
\usepackage{listings}
\usepackage{amsmath}
\usepackage{xspace}
\usepackage{tikz}

\usepackage{pifont}
\newcommand{\cmark}{\ding{51}}%
\newcommand{\xmark}{\ding{55}}%
\usepackage{url}
\usepackage{multirow}
\usepackage{caption}
\usepackage{subcaption}
\usepackage{amssymb}
\DeclareUnicodeCharacter{2061}{}
\usepackage{epstopdf}   
\usepackage{epsfig}
\usepackage{graphicx}
\usepackage{comment}
\usepackage{amsthm}

\usepackage{enumerate}
\usepackage[utf8]{inputenc}
\usepackage{lettrine}
\usepackage{hyperref}
\usepackage{array}
\usepackage{stfloats}
\usepackage{blindtext}
\usepackage{fancyhdr}
\usepackage{graphicx}
\usepackage{verbatim}
\usepackage{xcolor}
\usepackage{url}
\usepackage{comment}
\usepackage{lettrine}
\usepackage{cite}
\usepackage[normalem]{ulem}
\usepackage{pifont}
\begin{document}

\title{A Joint Time and Energy-Efficient Federated Learning-based Computation Offloading Method for Mobile Edge Computing}
\author{
Anwesha Mukherjee and Rajkumar Buyya, \IEEEmembership{Fellow, IEEE}
\thanks{A. Mukherjee is with the Department of Computer Science, Mahishadal Raj College, Mahishadal, West Bengal, 721628, India, and a Research Visitor in the Cloud Computing and Distributed Systems (CLOUDS) Laboratory, School of Computing and Information Systems, University of Melbourne, Australia. (e-mail: anweshamukherjee2011@gmail.com).}
\thanks{R. Buyya is with the Cloud Computing and Distributed Systems (CLOUDS) Laboratory, School of Computing and Information Systems, University of Melbourne, Australia. (e-mail: rbuyya@unimelb.edu.au).}
}
\maketitle

\begin{abstract}
Computation offloading at lower time and lower energy consumption is crucial for resource limited mobile devices. This paper proposes an offloading decision-making model using federated learning. Based on the task type and the user input, the proposed decision-making model predicts whether the task is computationally intensive or not. If the predicted result is \textit{computationally intensive}, then based on the network parameters the proposed decision-making model predicts \textit{whether to offload or locally execute} the task. According to the predicted result the task is either locally executed or offloaded to the edge server. The proposed method is implemented in a real-time environment, and the experimental results show that the proposed method has achieved above 90\% prediction accuracy in offloading decision-making. The experimental results also present that the proposed offloading method reduces the response time and energy consumption of the user device by  $\sim$11-31\% for computationally intensive tasks. A partial computation offloading method for federated learning is also proposed and implemented in this paper, where the devices which are unable to analyse the huge number of data samples, offload a part of their local datasets to the edge server. For secure data transmission, cryptography is used. The experimental results present that using encryption and decryption the total time is increased by only 0.05-0.16\%. The results also present that the proposed partial computation offloading method for federated learning has achieved a prediction accuracy of above 98\% for the global model. 
\end{abstract}

\begin{IEEEkeywords}
Federated learning, Offloading, Training Time, Encryption, Energy consumption.
\end{IEEEkeywords}

\section{Introduction}
The energy and latency-aware computation offloading has gained a significant research interest in the last few years with the rapid growth in wireless technology and the growing popularity of Internet of Things (IoT). The IoT devices and the mobile devices including laptops, tablets, smartphones, have limited battery life and computational resources. In such a case, execution of computationally intensive tasks may not be feasible inside these mobile devices. However, if the device is able to execute the tasks then also it may take more time than offloading the task to the edge server or cloud. On the other hand, offloading of all the tasks may not be fruitful from the perspective of latency and energy consumption. For the tasks with less computations, offloading can consume more time than local execution. Hence, the decision of \textit{offload or local execution} is important. 
\par
Nowadays, the mobile devices have various types of recommendation applications, where model training is important. There are many applications which use reinforcement learning, machine learning, or deep learning algorithms to train the model. Further, local model training may not be always possible for the resource-constrained mobile devices. However, the user's data is confidential and the user may not like to share the data with the server. Also, the large amount of data transmission to the server for analysis creates a huge overhead on the server as well as highly increase the network traffic. Federated learning (FL) \cite{nguyen2021federated, wu2024fedcache, yao2024ferrari} is an emerging technology where the devices locally train models using their local datasets, and exchange model updates with the server. Finally, a global model is developed, and each device that serves as a client has a personalized model and the global model. The use of FL protects data privacy as no data sharing takes place, and through a collaborative learning a global model with high prediction accuracy can be obtained. However, the device may not able to analyse a large dataset using deep learning algorithms. In that case, the dataset is partitioned into two parts: one part is used for local model training, and the other part is offloaded to the server. However, during data offloading to the server, protection of data privacy is vital.     
\subsection{Motivation and Contributions}
A mobile device has a computational task to execute. Now, the decision regarding whether to locally execute it or offload it to the edge server, is crucial with respect to the response time and energy consumption. Hence, a decision-making model is required to take the decision \textit{whether to offload or locally execute}. The data samples used for training may not be very high for a mobile device, and the network parameters' values also change. In such a case, a deep learning model with comparatively small number of samples may not provide prediction with high accuracy. Further, the users' data are different, and the local model weights are different. Thus, a collaborative training can provide a better prediction model so that the response time for task execution and energy consumption of the user device during the period can be minimal. The first motivation of this work to develop a collaborative decision-making model with high prediction accuracy to decide \textit{whether to offload or locally execute}. \par 
The use of FL for collaborative model training with privacy protection has gained popularity for various IoT applications. In centralized federated learning (CFL), the devices working as clients train local models with their own datasets, and exchange model updates with the server to build a global model \cite{nguyen2021federated}. In decentralized federated learning (DFL), the devices form a collaborative network among themselves to exchange model updates for developing a generalized model \cite{nguyen2021federated}. However, the mobile devices may not be always able to locally train deep learning model using a huge number of data samples. In such a case, conventional CFL and DFL models may not be feasible. The second motivation of this work is to propose a secure partial computation offloading framework for FL to deal with this issue.  \par
To address these issues, our paper makes the following contributions:
\begin{itemize}
    \item To address the first motivation, a CFL method is proposed for decision-making regarding \textit{whether to offload or locally execute}. The proposed model is named as \textit{\underline{FL}-based Offloading \underline{Dec}ision-Making Model} (\textit{FLDec}). At first \textit{FLDec} is used to decide whether a task is computationally intensive or not, based on the requested computation and user input. Here, as the underlying data analysis model Multi-layer Perceptron (MLP) is used, which is suitable for nonlinear function classification tasks. If the prediction result is \textit{computationally intensive}, then the \textit{FLDec} is used to decide whether to offload the task or locally execute it, based on the network parameters such as uplink and downlink traffic, network throughput, etc. Here, Long Short-Term Memory (LSTM) network is used as the underlying data analysis model, which is suitable for sequential data analysis. As the network parameters like throughput, uplink and downlink traffic are dynamic and change over time, we use LSTM in this case as the underlying model to capture the sequential pattern. 
    \item To address the second motivation, a secure partial computation offloading method for FL is proposed and implemented, which is referred to as \textit{\underline{Fed}erated \underline{Off}loading} (\textit{FedOff}). The user devices have large data samples, which are split into two parts. One part is used as the local dataset to train the local model, and the other part is offloaded to the edge server after encryption. The devices train their local models and exchange model updates with the edge server. The edge server after receiving data from the devices, trains a model with the received dataset after decryption, and stores the model update. After receiving model updates from the devices i.e. clients, the server performs aggregation considering the received updates as well as the stored update. The updated global model is then shared with the devices. In this work, we use symmetric key cryptography for data encryption during transmission to the edge server. However, the keys of individual devices are different. 
\end{itemize}
The rest of the paper is organized as follows. Section \ref{rel} presents the existing works on offloading and FL. Section \ref{pro} demonstrates the proposed offloading method. Section \ref{perf} evaluates the performance of the proposed approach. Finally, Section \ref{con} concludes the paper.

\section{Related Work}
\label{rel}
The existing research works on computation offloading, FL, and federated offloading are briefly discussed in this section. 
\par
The computation offloading is a significant research area of mobile cloud computing (MCC). Service delay minimization \cite{chen2023service} is one of the primary issues for mobile devices. The computation offloading from mobile device to the remote cloud suffered from delays, connectivity interruption, etc. \cite{mukherjee2022mcg}. The use of edge computing (EC) overcomes this problem by bringing resources at the network edge. The IoT/mobile devices are connected with the edge server, and the edge server is connected with the cloud \cite{mukherjee2021introduction}. The mobile devices offload the resource exhaustive computations to the edge server. The edge server executes the computation and sends the result to the mobile device, or forwards the request to the cloud server. The cloud server after executing the computation sends the result. The computation offloading in mobile edge computing (MEC) was discussed in \cite{feng2022computation}. The computation offloading in EC networks was illustrated in \cite{sadatdiynov2023review} along with a discussion on optimization methods. A joint task offloading approach with resource allocation was proposed for MEC environment in \cite{jiang2022joint}. In \cite{mukherjee2022mcg}, user mobility-based computation offloading method was proposed using game theory. In \cite{taheri2023machine}, the use of machine learning (ML) in computation offloading was reviewed. The reinforcement learning (RL) also gained popularity in computation offloading. In \cite{zabihi2023reinforcement}, the use of RL in computation offloading was discussed in detail. The use of deep RL for task offloading was discussed in \cite{tang2020deep}. A deep RL-based method for action space recursive decomposition was proposed in \cite{ho2020joint}. FL with edge computing has gained popularity in recent years \cite{lim2020federated, ye2020edgefed}. In \cite{bera2024flag}, an edge-based FL was proposed where the edge servers train their local models and exchange updates with the cloud to build the global model. However, in MEC-based FL, the mobile devices can also train their local models with own datasets and exchange model updates with the edge server that is connected with the cloud. The edge server performs aggregation in that case. The edge servers can also exchange their model updates with the cloud for developing a global model, and in that case cloud server acts as the aggregator. In \cite{liu2023federated}, FL with deep RL was used for computation offloading. Bidirectional LSTM (Bi-LSTM)-based FL was proposed for offloading decision-making and resource allocation in \cite{zhang2023bilstm}. In \cite{han2019federated}, FL was used for computation offloading in Edge-IoT systems. In \cite{tang2024federated}, FL-based task offloading was proposed based on multi-feature matching. To mitigate the straggler effect of FL, an offloading method was proposed in \cite{ji2021computation}. To deal with straggler effect and accelerate the local training in resource-constrained devices, offloading in FL was discussed in \cite{wu2022fedadapt}.    
\par
\textit{Limitations of existing approaches and comparison with proposed work:} In the existing FL-based offloading approaches, either the offloading decision-making using FL or partial computation offloading in FL, was explored. In our work, we propose a computation offloading decision-making framework using FL, as well as a partial computation offloading method for FL. Unlike the related works on offloading decision-making \cite{jiang2022joint, tang2020deep, ho2020joint, zhang2023bilstm, tang2024federated}, we have not only measured the time and energy consumption for the proposed methods but also the accuracy and model loss for the proposed approaches. In \cite{liu2023federated} and \cite{han2019federated}, the authors though measured the model loss, the energy consumption was not measured. In comparison with the existing partial computation offloading methods \cite{ji2021computation, wu2022fedadapt}, the uniqueness of the proposed work is the use of cryptography for secure data transmission to the server. The comparison of the proposed and existing offloading approaches is presented in Table \ref{tab:comoff}. From the table we observe that the proposed approach is unique compared to the existing schemes.  

\begin{table*}
\caption{\centering{Comparison of proposed work with existing offloading methods}}
    \centering
    \begin{tabular}{c c c c c c c}
        \hline
         Work & Offloading decision-- & Partial computation & Task details & Accuracy/ Loss & Time/ Delay & Energy consumption\\
        &  making using FL  & offloading for FL & are provided &is measured & is measured& is measured\\
        \hline
        Jiang et al. \cite{jiang2022joint} & \xmark & \xmark & \cmark & \xmark & \cmark & \cmark\\
        Tang et al. \cite{tang2020deep} & \xmark & \xmark & \xmark & \xmark & \cmark & \xmark\\
        Ho et al. \cite{ho2020joint} & \xmark & \xmark & \cmark & \xmark & \cmark & \cmark\\
    Liu et al. \cite{liu2023federated} & \cmark & \xmark & \xmark & \cmark & \cmark & \xmark\\
   Zhang et al. \cite{zhang2023bilstm} & \cmark & \xmark & \xmark & \xmark & \cmark & \cmark\\
   Han et al. \cite{han2019federated} & \cmark & \xmark & \xmark & \cmark & \xmark & \xmark\\
   Tang et al. \cite{tang2024federated} & \cmark & \xmark & \cmark & \xmark & \cmark & \cmark\\
  Ji et al. \cite{ji2021computation} & \xmark & \cmark & \cmark & \cmark & \cmark & \xmark\\
   Wu et al. \cite{wu2022fedadapt} & \xmark & \cmark & \cmark & \cmark & \cmark & \xmark\\
  Proposed work & \cmark & \cmark & \cmark & \cmark & \cmark & \cmark\\
       \hline
    \end{tabular}
    \label{tab:comoff}
\end{table*}

\section{Proposed Offloading Methods}
\label{pro}
The proposed approach is divided into two parts. The first part is on decision-making regarding offloading. In this case, we have used FL to build a decision-making model regarding whether to offload or not. The second part is on partial computation offloading in FL. In the second part, we propose a secure partial computation offloading framework for FL for resource limited mobile devices. 
\par
\begin{figure}
\centering
    \includegraphics[width=0.99\linewidth, height=2.2in]{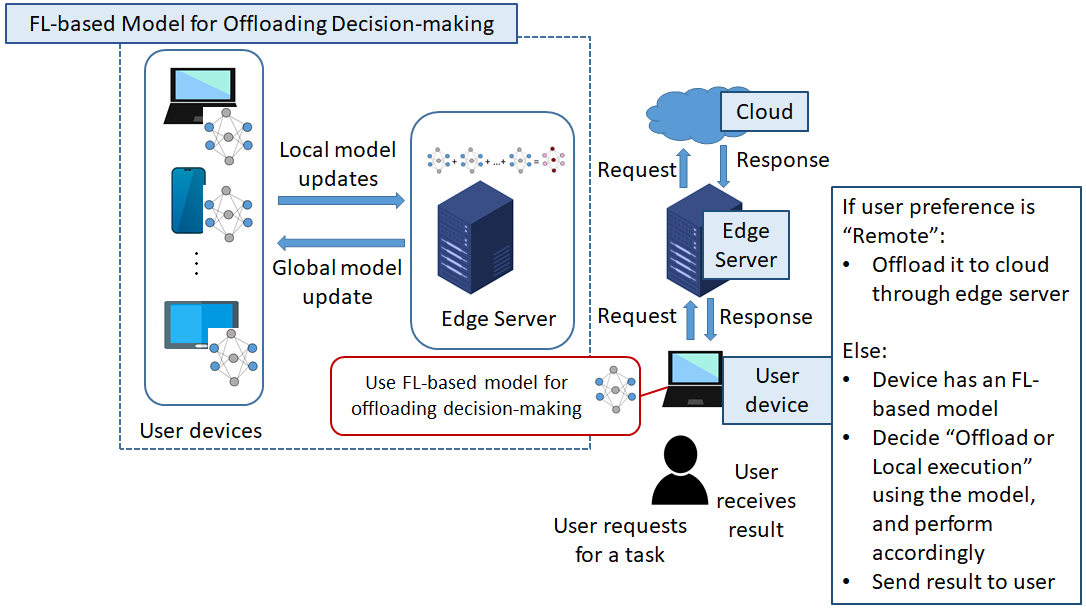}
    \caption{\centering{System model of the proposed offloading decision-making process using FL}}
    \label{arch1}
\end{figure}

\subsection{Proposed Offloading Decision-making Method}
We propose a decision-making method regarding whether to offload a task or not, based on its computational intensiveness, network parameters, etc. The mathematical notations used in our approach FLDec are summarized in Table \ref{tab:sym}. 
\begin{table} []
    \centering
    \caption{Mathematical notations used in FLDec}
    \begin{tabular}{c|c}
    \hline
       Notation  &  Definition\\
       \hline
       $\mathcal{K}$ & Set of connected devices\\
       $\mathcal{D}_k$  & Local dataset of a device $k$\\
      $B$ & Batchsize into which $\mathcal{D}_k$ is split\\
      $\mathcal{E}$ & Number of epochs\\
      $\mathcal{R}$ & Number of rounds\\
      $\mathcal{M}_{f}$ & Final global model\\
      $\mathcal{M}_k$ & Local model of device $k$\\
      $\alpha_r$ & Fractions of devices participating in round $r$\\
      $N_k$ & Maximum number of devices for participating in FL\\
       $\mathcal{M}_{k_{exh}}$ & Local model of a device $k$ to predict \\& whether a task is computationally intensive or not\\
      $\mathcal{M}_{k_{off}}$ & Local model of a device $k$ to predict\\ &whether to offload or not\\
     \hline
    \end{tabular}
    \label{tab:sym}
\end{table}

The problem statement and the proposed method are discussed as follows. 
\subsubsection{Problem statement}
A device has a computational task $C$ that is to be executed locally or remotely. 
\begin{itemize}
    \item \textit{Objective 1:} Decide whether $C$ is computationally intensive or not. 
    \item \textit{Objective 2:} If it is computationally intensive, then decide whether to offload $C$ or not. 
\end{itemize}
\subsubsection{Proposed framework} 
To address the first and second objectives, we use the FL-based model. The steps of developing a decision-making model using FL is presented in Algorithm \ref{algo_1}. Here, the clients are the participating user devices, and the server is the edge server. 

\begin{algorithm} 
\caption{FL-based model for decision-making}
\label{algo_1}
\small
\KwIn{$\mathcal{D}_k$, $\mathcal{K}$, $\mathcal{R}$, $B$, $\mathcal{E}$, $N_k$}
\KwOut{Updated Global model ($\mathcal{M}_f$), Local model ($\mathcal{M}_k$)}
\BlankLine
\SetKwFunction{FAlgoName}{Decision-making Model}
\SetKwProg{An}{}{}{}
\An{\FAlgoName$(\mathcal{D}_k, \mathcal{K}, \mathcal{R})$:} 
{
    $j \gets 0$\\
    \While{($j < N_k$)}
    {
    $listen()$\\
    $establishConnection()$\\
    $j=j+1$
    }
    $\mathcal{K} \gets ConnectedDevices()$\\
    start $\text{Client-side process()}$ and $\text{Server-side process()}$
}
\An{$\text{Client-side process()}$:} 
{
    $\mathcal{D}_k \; \gets \; preprocess(\mathcal{D}_k)$\\
    $Accuracy \gets 0$\\
    \While{($connectedtoServer()==TRUE$)}
    {
    $\mathcal{M}_{up} \gets get(\mathcal{M})$ \\
    $\mathcal{D}_{k_{train}}, \mathcal{D}_{k_{test}} \gets split(\mathcal{D}_k)$\\
     $\mathcal{M}_{up}.fit(\mathcal{D}_{k_{train}}, \mathcal{E}, B)$\\
    $Accuracy_{up} \gets accuracy\_score(\mathcal{M}_{up}, \mathcal{D}_{k_{test}})$\\
    \If{($Accuracy < Accuracy_{up}$)} {$\mathcal{M}_{k} \gets \mathcal{M}_{up}$}
    $send(\mathcal{M}_{k})$
    }
    $save(\mathcal{M}_{k})$
}

\An{$\text{Server-side process():}$} 
{
    $\mathcal{M}_{in} \gets initmodel()$\\
    $\mathcal{M} \gets \mathcal{M}_{in}$\\
    \For{$r=1 \; to \; \mathcal{R}$}
    {
    initialize $Model_{all}=[]$\\
    $\mathcal{K}_r \gets Subset(\mathcal{K}, max(\alpha_r*|\mathcal{K}|,1),``random")$\\
   \For{$k=1 \; to \; |\mathcal{K}_r|$}
   {
   $receive(\mathcal{M}_k)$\\
   $Model_{all}.append(\mathcal{M}_k)$
   }
   $w \gets \sum_{k=1}^{|\mathcal{K}_r|} getweights(\mathcal{M}_k)/|\mathcal{K}_r|$\\
   $\mathcal{M}.setweights(w)$\\
   $send(\mathcal{M})$
   }
   $\mathcal{M}_f \gets \mathcal{M}$\\
   $save(\mathcal{M}_f)$\\
   $releaseConnectedDevices()$
}
\end{algorithm}

For the first objective, we use MLP as the underlying data analysis model. Based on the requested computational task and the respective input, it is predicted whether the task would be computationally intensive or not. For a computational task, the input plays an important role. For example, multiplication of two 3x3 matrices takes much less time to compute than the multiplication of two 300x300 matrices. Thus, based on the input as well as the task type (matrix operation, sorting, searching, etc.), it is decided whether it can be computationally intensive or not. To address the second objective, we use LSTM as the underlying model. Based on some network parameters such as network throughput, latency, uplink and downlink traffic, etc., it is decided whether to offload the task or not. If the decision is \textit{offload}, $C$ is offloaded to the edge server. If the edge server is unable to offload $C$, then it forwards the request to the cloud. The cloud server executes the task and sends the result to the device. The proposed offloading method is presented in Algorithm \ref{algo_2}. The proposed method is pictorially depicted in Fig. \ref{arch1}.

\begin{algorithm} 
\caption{Proposed offloading algorithm}
\label{algo_2}
\small
\KwIn{Computational Task ($C$), Model to check computational intensiveness ($\mathcal{M}_{k_{exh}}$), Model to check offload or local execution ($\mathcal{M}_{k_{off}}$)}
\KwOut{Result ($R$)}
\BlankLine
\SetKwFunction{FAlgoName}{Offloading method}
\SetKwProg{An}{}{}{}
\An{\FAlgoName$(C)$:} 
{
$category(C) \gets \mathcal{M}_{k_{exh}}.predict(inputfeatures_{exh})$\\
    \If{$category(C)$ is $NotIntensive$}
    {
    \If{($userPreference==``RemoteAccess"$)}
    {$R \gets offloadtoCloudthroughEdgeServer(C)$}
    \Else{$R \gets executeLocally(C)$}
    }
    \Else{
    $decision \gets \mathcal{M}_{k_{off}}.predict(inputfeatures_{off})$\\
    \If{($decision==``Offload"$)}
    {$R \gets offloadtoEdgeServer(C)$}
    \Else{$R \gets executeLocally(C)$}
    }
}
\end{algorithm}

\subsubsection{Computational complexity} The time complexity of the FL process depends on the time complexity of model initialization, local model training, exchange of model updates, and aggregation. The time complexity of model initialization is $O(1)$. The computational complexity of model initialization is given as $O(w_{\mathcal{M}_{in}})$, where $w_{\mathcal{M}_{in}}$ denotes the weights of the initial model. The time complexity of local model training is given as $O(\mathcal{R} \cdot \mathcal{E} \cdot (\mathcal{D}_k/B) \cdot w_k)$, where $w_k$ denotes model weights for client $k$. The computational complexity of local model training is given as $O(\mathcal{R} \cdot \mathcal{E} \cdot (\mathcal{D}_k/B) \cdot w_k \cdot |\mathcal{K}|)$. The time complexity for model aggregation is given as $O(\mathcal{R} \cdot |\mathcal{K}| \cdot w_k)$. The computational complexity for model aggregation is also given as $O(\mathcal{R} \cdot |\mathcal{K}| \cdot w_k)$. The time complexity and computational complexity for model updates exchange both are given as $O(w_{\mathcal{M}} + \mathcal{|K|} \cdot w_k)$, where $w_k$ denotes the model weights of client $k$ and $w_\mathcal{M}$ denotes model weights of the server. 
\subsection{Proposed Federated Offloading Method}
In this section, we first discuss the problem statement, and then propose a partial computation offloading method for FL, which is referred to as federated offloading. In the proposed method, the user devices are the clients and the server is the edge server. The mathematical notations used in our approach FedOff are summarized in Table \ref{tab:sym1}. 
\subsubsection{Problem statement} 
Let the model of an application $A$ needs to be developed using FL, each device has a respective local dataset $\mathcal{D}_{k_A}$, and the number of rounds is $\mathcal{R}_A$. In FL, each of the participating device needs to locally train the model with its local dataset. However, model training with a large number of data samples may not be feasible for all the devices. In that case, the devices put a portion of their datasets to the server. The server trains the model with the offloaded datasets and stores the model. During aggregation, the server considers this trained model also along with the received model updates from the devices. However, how much portion of data will be offloaded that is significant. Further, data security during offloading to the server is another issue. The objective is to propose a secure federated offloading method with good prediction accuracy and minimal loss.\par

\begin{table} []
    \centering
    \caption{Mathematical notations used in FedOff}
    \begin{tabular}{c|c}
    \hline
       Notation  &  Definition\\
       \hline
       $\mathcal{K}$ & Set of connected devices\\
       $\mathcal{D}_{k_A}$  & Local dataset of a device $k$ for application $A$\\
      $B_A$ & Batchsize into which $\mathcal{D}_{k_A}$ is split\\
      $\mathcal{E}$ & Number of epochs\\
      $\mathcal{R}_A$ & Number of rounds in federated offloading\\
      $\mathcal{M}_{f_A}$ & Final global model for application $A$\\
      $\mathcal{M}_{k_A}$ & Local model of device $k$ for application $A$\\
      $\alpha_r$ & Fractions of devices participating in round $r$\\
      $N_k$ & Maximum number of devices for participating in FL\\
      $T_{total}$ & Total time consumption in FedOff\\
      $T_{train}$ & Total time consumption for training in FedOff\\
      $T_{init}$ & Time consumption for model initialization\\
      $T_{tr}$ & Time consumption for data transmission\\
      &from clients to server\\
      $T_{loc}$ & Time consumption for local model training\\
      $T_{exm}$ & Time consumption for model updates exchange\\
      $T_{ser}$ & Time consumption for model training
      inside \\
      & the server using received local datasets\\
      $T_{agg}$ & Time consumption for aggregation of model updates\\
      $T_{crypt}$ & Time consumption for data encryption and decryption\\
     \hline
    \end{tabular}
    \label{tab:sym1}
\end{table}

\begin{figure}
\centering
    \includegraphics[width=0.995\linewidth, height=2.5in]{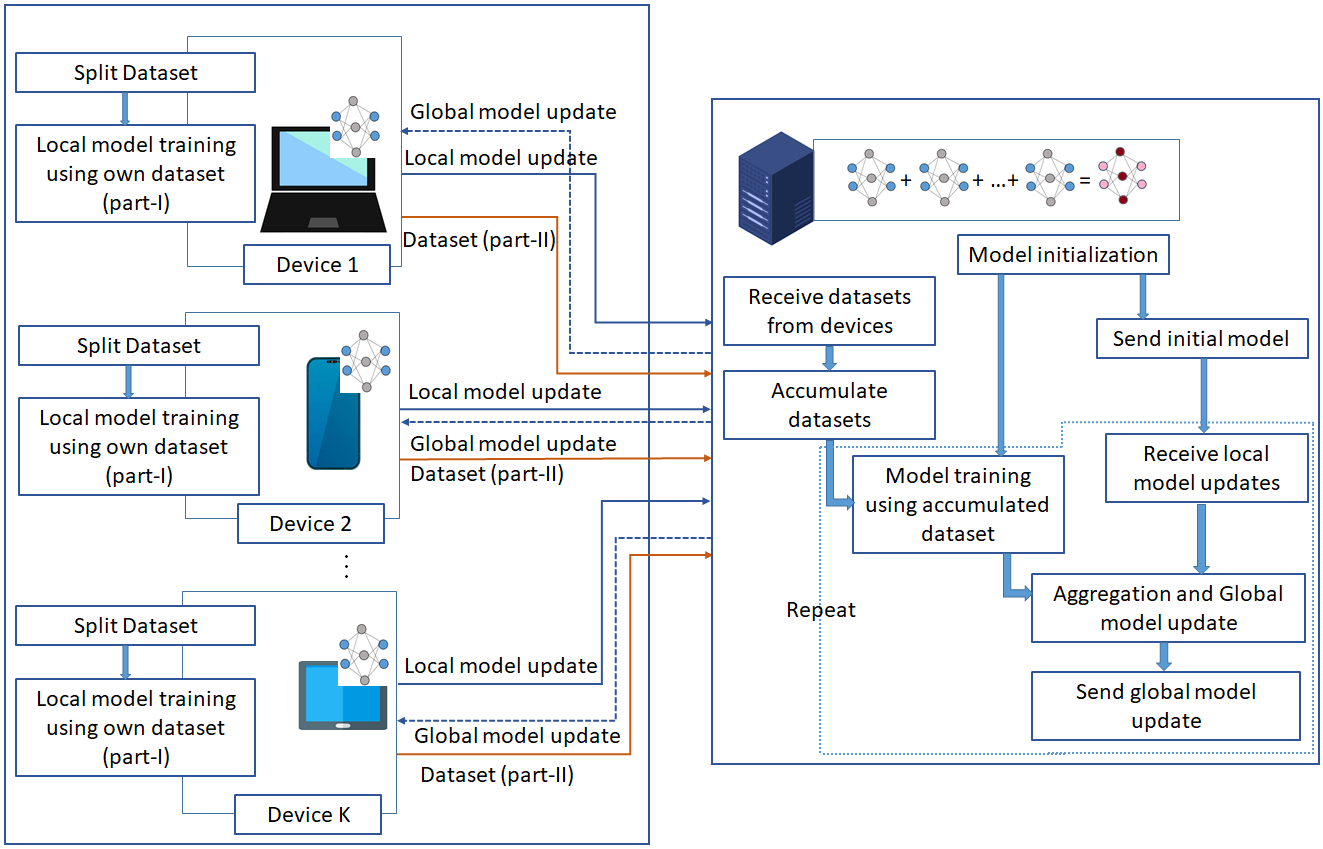}
    \caption{\centering{The proposed federated offloading method}}
    \label{arch2}
\end{figure}

\subsubsection{Proposed framework}
In the proposed federated offloading method, each participating user device $k$ splits its local dataset into two parts: one part will be processed locally ($D_{k_{loc}}$) and the other part will be sent to the edge server ($D_{k_{ser}}$). After splitting the dataset, the client sends the second part ($D_{k_{ser}}$) to the server. The server receives $D_{k_{ser}}$ from each participating device $k$, accumulates the collected data, and fits its model with the dataset. Each device $k$ trains its model with local dataset ($D_{k_{loc}}$) and sends the model update to the server. The server after receiving model updates from the clients performs aggregation along with the model update for the offloaded data. The process of model update is repeated for the number of rounds $\mathcal{R}_A$. The proposed federated offloading method is presented in Fig. \ref{arch2}. During data transmission from the clients to the server, encryption is used for protecting the data from unauthorized access. 

\begin{algorithm} 
\caption{Federated offloading method}
\label{algo_3}
\small
\KwIn{$\mathcal{D}_{k_A}$, $\mathcal{K}$, $\mathcal{R}_A$, $B_A$, $\mathcal{E}$, $N_k$}
\KwOut{Updated Global model ($\mathcal{M}_{f_A}$), Local model ($\mathcal{M}_{k_A}$)}
\BlankLine
\SetKwFunction{FAlgoName}{Update Model}
\SetKwProg{An}{}{}{}
\An{\FAlgoName$(\mathcal{D}_{k_A}, \mathcal{K}, \mathcal{R}_A)$:} 
{
    $j \gets 0$\\
    \While{($j < N_k$)}
    {
    $listen()$\\
    $establishConnection()$\\
    $j=j+1$
    }
    $\mathcal{K} \gets ConnectedDevices()$\\
    start $\text{Client-side process()}$ and $\text{Server-side process()}$
}
\An{$\text{Client-side process()}$:} 
{
    $\mathcal{D}_{k_A} \; \gets \; preprocess(\mathcal{D}_{k_A})$\\
    $\mathcal{D}_{k_{loc}}, \mathcal{D}_{k_{ser}} \gets split(\mathcal{D}_{k_A})$\\
    $Accuracy \gets 0$\\
    $t \gets 1$\\
    \While{($connectedtoServer()==TRUE$)}
    {
    \If{($t==1$)}{$\mathcal{D}_{k_{ser}} \gets encrypt(\mathcal{D}_{k_{ser}})$\\$sendtoserver(\mathcal{D}_{k_{ser}})$\\ $t \gets t+1$}
    $\mathcal{M}_{up} \gets get(\mathcal{M}_A)$ \\
    $\mathcal{D}_{k_{train}}, \mathcal{D}_{k_{test}} \gets split(\mathcal{D}_{k_{loc}})$\\
     $\mathcal{M}_{up}.fit(\mathcal{D}_{k_{train}}, \mathcal{E}, B_A)$\\
    $Accuracy_{up} \gets accuracy\_score(\mathcal{M}_{up}, \mathcal{D}_{k_{test}})$\\
    \If{($Accuracy < Accuracy_{up}$)} {$\mathcal{M}_{k_A} \gets \mathcal{M}_{up}$}
    $send(\mathcal{M}_{k_A})$
    }
    $save(\mathcal{M}_{k_A})$
}
\An{$\text{Server-side process():}$} 
{
    $\mathcal{M}_{in} \gets initmodel()$\\
    $\mathcal{M}_A \gets \mathcal{M}_{in}$\\
    initialize $\mathcal{D}_{loc}=[]$\\
    \For{$k=1 \; to \; \mathcal{|K|}$}
    {
    $receive(\mathcal{D}_{k_{ser}})$\\
    $\mathcal{D}_{k_{ser}} \gets decrypt(\mathcal{D}_{k_{ser}})$\\
    $\mathcal{D}_{loc}.append(\mathcal{D}_{k_{ser}})$
    }
    $\mathcal{D}_{l_{train}}, \mathcal{D}_{l_{test}} \gets split(\mathcal{D}_{loc})$\\
    $Accuracy_{l} \gets 0$\\
    \For{$r=1 \; to \; \mathcal{R}_A$}
    {
    initialize $Model_{all}=[]$\\
    $\mathcal{K}_r \gets Subset(\mathcal{K}, max(\alpha_r*|\mathcal{K}|,1),``random")$\\
   \For{$k=1 \; to \; |\mathcal{K}_r|$}
   {
   $collect(\mathcal{M}_{k_A})$\\
   $Model_{all}.append(\mathcal{M}_{k_A})$
   }
    $\mathcal{M}_{loc_u} \gets \mathcal{M}_A$\\
     $\mathcal{M}_{loc_u}.fit(\mathcal{D}_{l_{train}}, \mathcal{E}, B_A)$\\
     $Accuracy_{loc} \gets accuracy\_score(\mathcal{M}_{loc_u}, \mathcal{D}_{l_{test}})$\\
     \If{($Accuracy_{l} < Accuracy_{loc}$)} {$\mathcal{M}_{loc} \gets \mathcal{M}_{loc_u}$}
     $Model_{all}.append(\mathcal{M}_{loc})$\\
     $N_{model} \gets length(Model_{all})$\\
   $w \gets \frac{1}{N_{model}}\sum_{\mathcal{M}_i \in Model_{all}} getweights(\mathcal{M}_i)$\\
   $\mathcal{M}_A.setweights(w)$\\
    $send(\mathcal{M}_A)$
   }
   $\mathcal{M}_{f_A} \gets \mathcal{M}_A$\\
   $save(\mathcal{M}_{f_A})$\\
   $releaseConnectedDevices()$
}
\end{algorithm}

\subsubsection{Computational complexity}
The time complexity of the proposed federated offloading method depends on the time complexity of model initialization, data transmission from clients to the server, local model training, exchange of model updates, model training at the server, and aggregation. The time complexity of model initialization is $O(1)$. The computational complexity of model initialization is given as $O(w_{\mathcal{M}_{in}})$, where $w_{\mathcal{M}_{in}}$ denotes the weights of the initial model. The time complexity and computational complexity of data transmission from clients to the server, both are given as $O(\mathcal{|K|} \cdot (\mathcal{D}_{k_{ser}}/S))$, where $S$ denotes the data transmission speed. The time complexity of local model training is given as $O(\mathcal{R}_A \cdot \mathcal{E} \cdot (\mathcal{D}_{k_{loc}}/B_A) \cdot w_{k_A})$, where $w_{k_A}$ denotes model weights for client $k$ for application $A$. The computational complexity of local model training is given as $O(\mathcal{R}_A \cdot \mathcal{E} \cdot (\mathcal{D}_{k_{loc}}/B_A) \cdot w_{k_A} \cdot \mathcal{|K|})$.
The time complexity and computational complexity for model updates exchange both are given as $O(w_{\mathcal{M}_A} + \mathcal{|K|} \cdot w_{k_A})$, where $w_{k_A}$ denotes the model weights of client $k$ and $w_{\mathcal{M}_A}$ denotes model weights of the server for application $A$.  
The time complexity of model training at the server is given as $O(\mathcal{R}_A \cdot \mathcal{E} \cdot (\mathcal{D}_{loc}/B_A) \cdot w_{\mathcal{M}_A})$, where $w_{\mathcal{M}_A}$ denotes the model weights of the server for application $A$. 
The time complexity and computational complexity for model aggregation both are given as $O(\mathcal{R}_A \cdot N_{model} \cdot w_i)$, where $w_i$ denotes model weights of $\mathcal{M}_i$ and $\mathcal{M}_i \in Model_{all}$. 

\subsubsection{Total time consumption}
The total time consumption in the proposed federated offloading method is determined as follows:
\begin{equation}
    T_{foff}=T_{train}+T_{crypt}
\end{equation}
where $T_{train}=T_{init}+T_{tr}+T_{loc}+T_{exm}+T_{ser}+T_{agg}$, where $T_{init}=O(1)$, $T_{tr}=O(\mathcal{|K|} \cdot (\mathcal{D}_{k_{ser}}/S))$, $T_{loc}=O(\mathcal{R}_A \cdot \mathcal{E} \cdot (\mathcal{D}_{k_{loc}}/B_A) \cdot w_{k_A})$, $T_{exm}=O(w_{\mathcal{M}_A} + \mathcal{|K|} \cdot w_{k_A})$, $T_{ser}=O(\mathcal{R}_A \cdot \mathcal{E} \cdot (\mathcal{D}_{loc}/B_A) \cdot w_{\mathcal{M}_A})$, and $T_{agg}=O(\mathcal{R}_A \cdot N_{model} \cdot w_i)$, where $w_i$ denotes model weights of $\mathcal{M}_i$ and $\mathcal{M}_i \in Model_{all}$.

\section{Performance Evaluation}
\label{perf}
In this section, firstly we analyse the performance of the proposed offloading decision-making method, and then analyse the performance of the proposed federated offloading framework. We have performed a real experimental analysis in the CLOUD lab, The University of Melbourne, to evaluate the performance of the proposed approaches. Python 3.8.10 has been used for implementation, and Tensorflow has been used for deep learning-based data analysis. For exchange of model updates in FL, \textit{MLSocket} has been used. 

\subsection{Performance of FLDec}
The proposed offloading decision-making model, \textit{FLDec}, is evaluated based on prediction accuracy, training time, training energy consumption, and the response time in task execution and energy consumption of the user device during that period considering different real-time offloading case studies. The time consumption is measured in seconds (s) and the energy consumption is measured in joule (J).
\subsubsection{Experimental setup}
For the experiment, we have provisioned four virtual machines from the RONIN cloud environment of Amazon AWS, to act as four user nodes. Each of the user nodes has 4GB memory. We also have provisioned two virtual machines from the RONIN cloud for using as the edge server and the cloud server instance. The edge server has 4GB memory and the processor is Intel(R) Xeon(R) Platinum 8259CL CPU @ 2.50GHz. The cloud server instance has 4GB memory and 2vCPUs. The number of rounds in FL has been set to 10, and $\alpha_r=1$. 

\subsubsection{Prediction accuracy, Training time, and Energy consumption of user device}
\begin{figure}
\centering
\includegraphics[width=0.995\linewidth, height=2.0in]{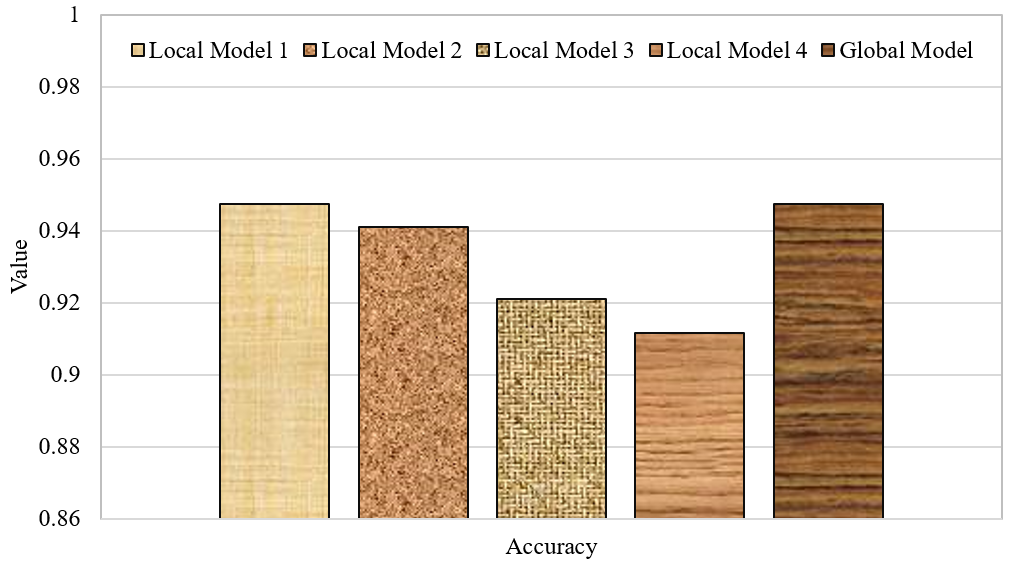}
    \caption{\centering{Performance of the local and global models for predicting the task is computational intensive or not}}
    \label{accoffmlp}
\end{figure}

\begin{figure}
\centering
\includegraphics[width=0.99\linewidth, height=1.8in]{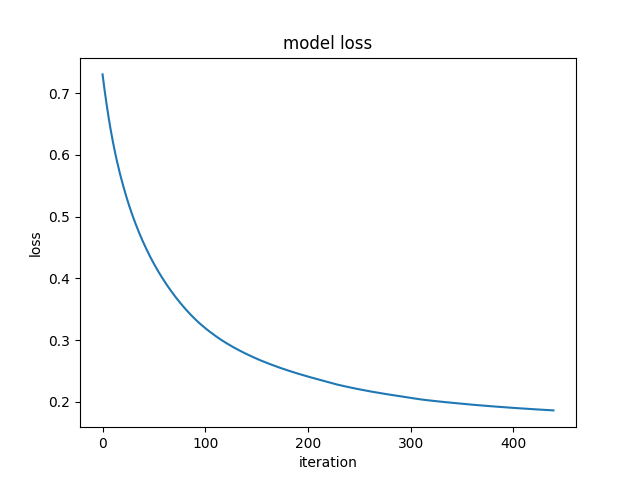}
    \caption{\centering{Global model loss while predicting the task is computational intensive or not}}
    \label{lossmlp}
\end{figure}

In the proposed framework, CFL is used. For predicting computational intensiveness of a task, we use MLP as the underlying model, and for offloading decision-making, we use LSTM as the underlying model. The parameter values used in MLP and LSTM are presented in Table \ref{tab:class}. We have used our own dataset\footnote{\url{https://github.com/AnuTuli/OffFed}} for predicting whether a task is computational intensiveness or not. We have executed several computational tasks such as basic calculator design, matrix multiplication, file creation, sorting, searching, etc., using different input values. Based on the outcomes, we developed the dataset. We have obtained the prediction accuracy of 94.74\% for the global model using the FL-based decision-making model, based on MLP. The results are presented in Fig. \ref{accoffmlp}. For the local models we have achieved an accuracy of 91.17-94.74\% after round three. The global model loss after round 10 is presented in Fig. \ref{lossmlp}. As we observe from the figure, the model converges when the loss becomes minimal. After predicting the computational intensiveness, we have used another dataset\footnote{\url{https://www.kaggle.com/datasets/ucimachinelearning/task-offloading-dataset/data}} that considers network parameters to predict whether to offload or locally execute, using the proposed FL-based decision-making model, based on LSTM. 
\begin{table}[]
\centering
\caption{\textcolor{black}{The values of the parameters in MLP and LSTM}}
\begin{tabular}{lll}
\hline
Classifier & Parameter & Value \\
\hline
LSTM       & \begin{tabular}[c]{@{}l@{}}Activation\\Optimizer\\Epochs\\Batch size\\\end{tabular}                    & \begin{tabular}[c]{@{}l@{}}softmax\\Adam\\10\\200\end{tabular}   \\
\hline
MLP       & \begin{tabular}[c]{@{}l@{}}Activation:\\ Maximum iteration:\\Solver:\\\end{tabular}                    & \begin{tabular}[c]{@{}l@{}}ReLU\\100\\Adam\end{tabular}   \\
\hline
\end{tabular}
\label{tab:class}
\end{table}

\begin{figure}
\centering
    \includegraphics[width=0.995\linewidth, height=2.0in]{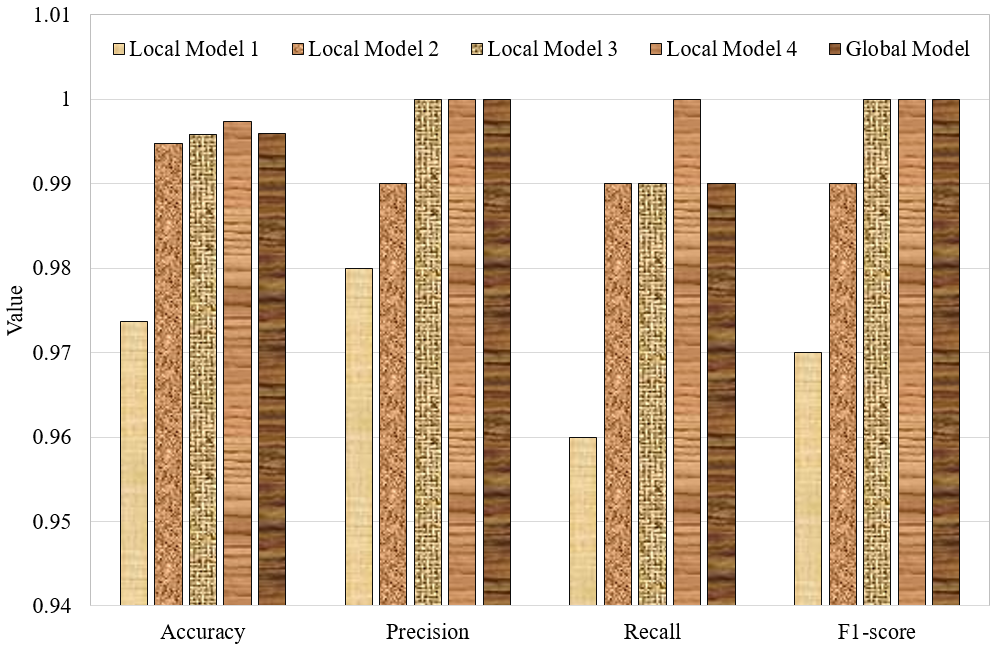}
    \caption{\centering{Performance of the local and global models for predicting whether to offload or locally execute the task}}
    \label{accoff}
\end{figure}

\begin{figure}
\centering
    \includegraphics[width=0.995\linewidth, height=2.2in]{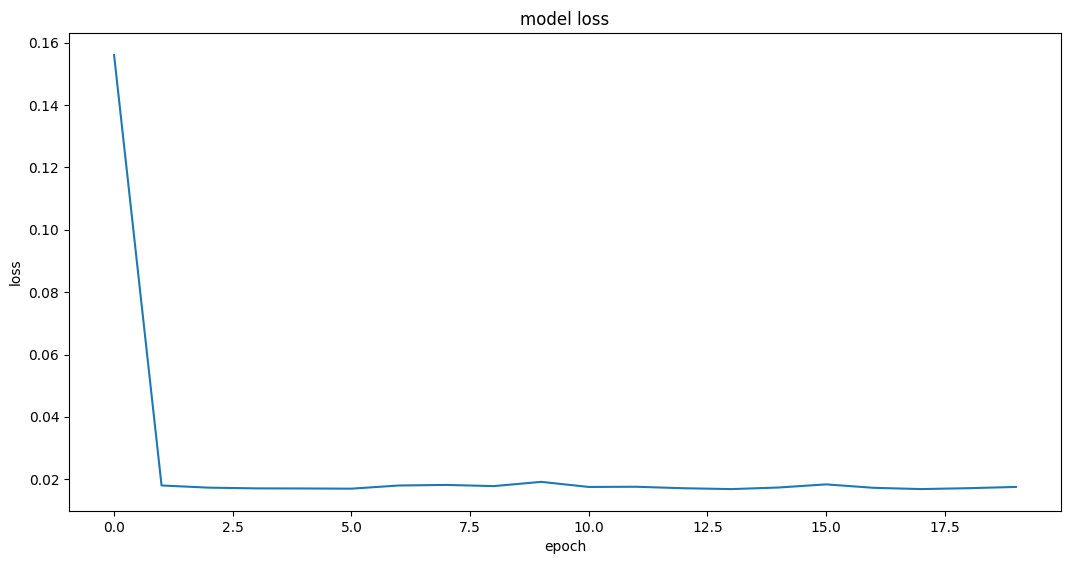}
    \caption{\centering{Global model loss while predicting whether to offload or locally execute the task}}
    \label{losslstm}
\end{figure}

\begin{figure*}
\centering
    \includegraphics[width=0.995\linewidth, height=1.2in]{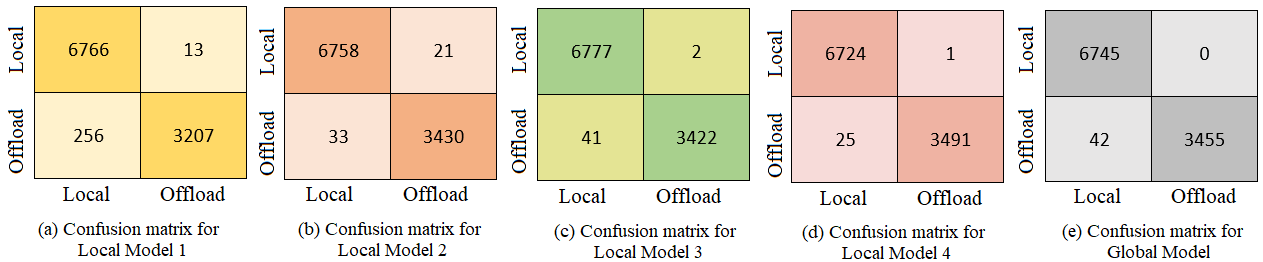}
    \caption{\centering{Confusion matrix for local and global models}}
    \label{conf}
\end{figure*}

The accuracy, precision, recall, and F1-score, achieved by the global model and the local models of the four participating devices after round 5 are presented in Fig. \ref{accoff}. As we observe, the global model has achieved an accuracy of $>$99\% and the local models have achieved $>$97\% accuracy. The precision, recall, and F1-score are $\geq$0.99 for the global model, and $\geq$0.96 for the local models. The global model loss after round 10 is presented in Fig. \ref{losslstm}, and we observe that the model converges when the loss becomes minimal. 
The confusion matrix of the global and local models are presented in Fig. \ref{conf}. Here, two classes are presented as the two decisions: (1) locally execute, (2) offload. 

\begin{figure} 
\centering
    \includegraphics[width=0.9\linewidth, height=2.2in]{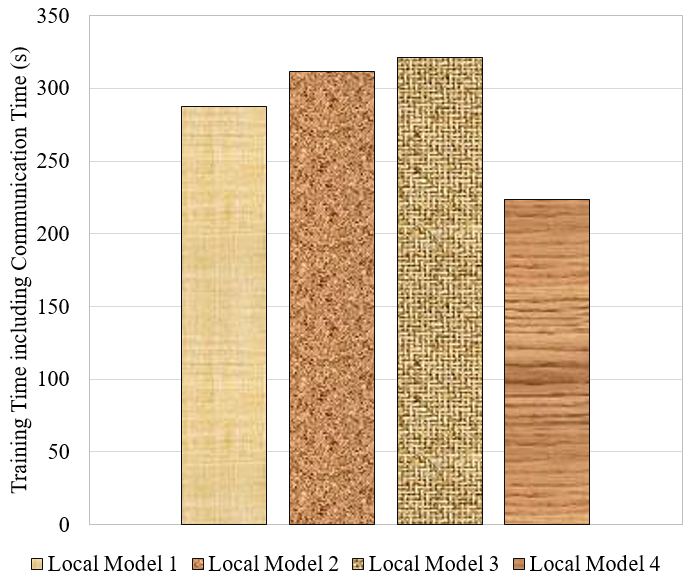}
    \caption{\centering{Training time including communication time for local models}}
    \label{trainoff}
\end{figure}

\begin{figure} 
\centering
    \includegraphics[width=0.9\linewidth, height=2.2in]{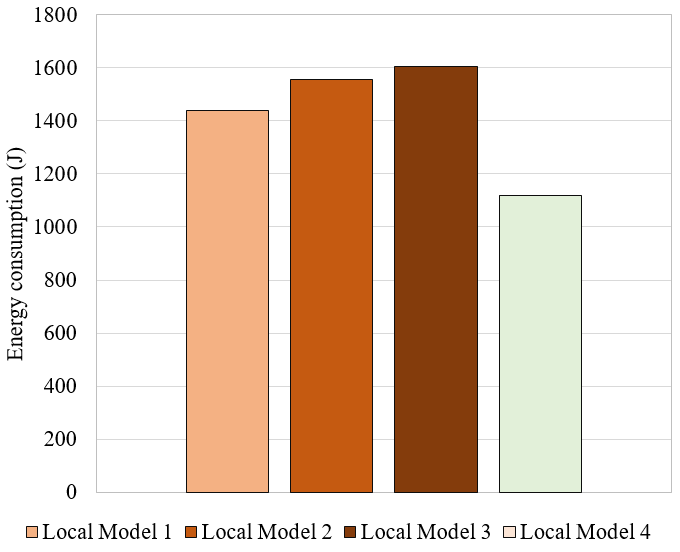}
    \caption{\centering{Energy consumption of the device during the training time including communication period}}
    \label{trainenoff}
\end{figure}

The total training time of the edge server including the communication time is 110.36s. The training time including the communication time for each of the participating devices are presented in Fig. \ref{trainoff}, and we observe that the time consumption is 200-350s for the local models. The energy consumption of the participating devices during the entire period are also measured and presented in Fig. \ref{trainenoff}, and we observe that the energy consumption of the device is 1000-1600J during the training periods. 
\par
\begin{figure} 
\centering
    \includegraphics[width=0.995\linewidth, height=2.2in]{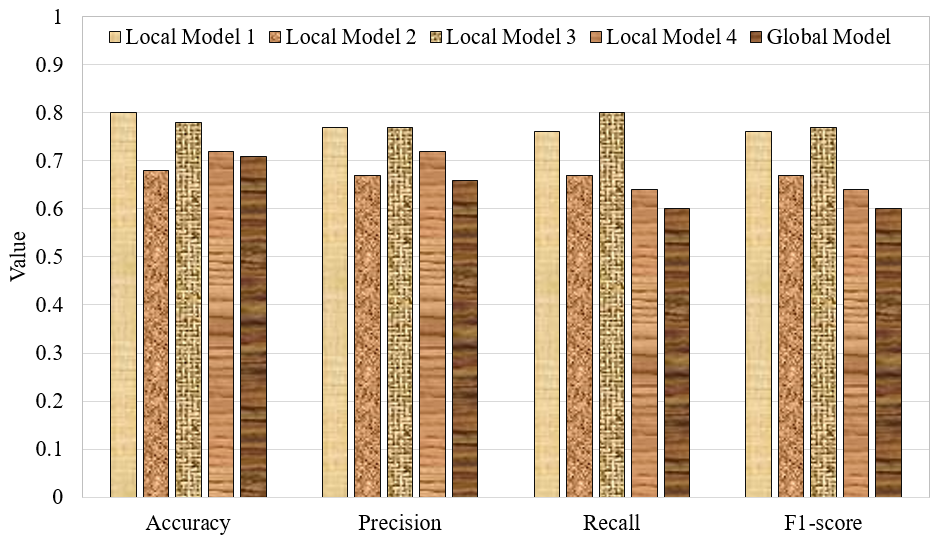}
    \caption{\centering{Performance of local and global models without using FL}}
    \label{acccomp}
\end{figure}

\textit{Comparison with decision-making without FL:} The accuracy, precision, recall, and F1-score achieved by the local and global models using LSTM without FL are also measured, and we have observed that the prediction accuracy is above 65\% for all the local models and above 70\% for the global model. The precision, recall, and F1-score for the all local models are above 0.6 and for the global model it is $\geq$0.6.  The results are presented in Fig. \ref{acccomp}. As we observe, the accuracy, precision, recall, and F1-score using FL are better for the global model as well as for all the local models.

\begin{figure} 
\centering
    \includegraphics[width=0.9\linewidth, height=2.0in]{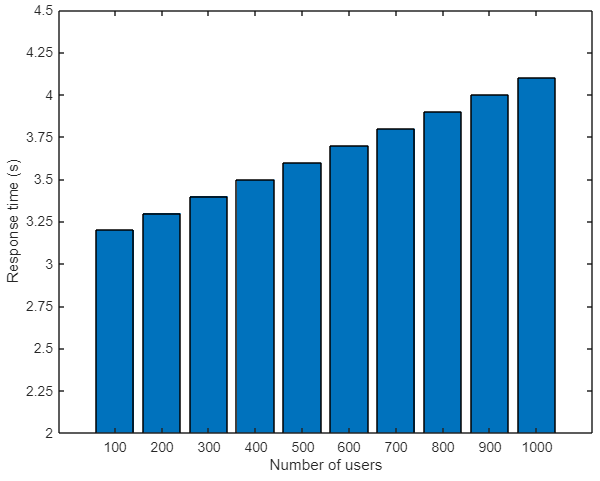}
    \caption{\centering{Average response time in FLDec}}
    \label{avgres}
\end{figure}

\begin{figure} 
\centering
    \includegraphics[width=0.9\linewidth, height=2.0in]{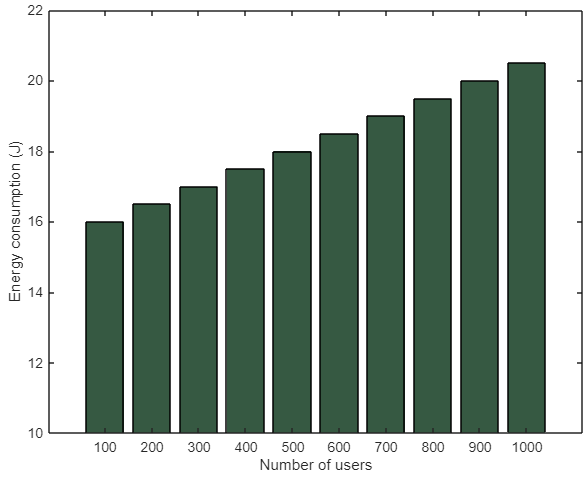}
    \caption{\centering{Average energy consumption of user device during task execution period in FLDec}}
    \label{avgener}
\end{figure}

\subsubsection{Task offloading, Response time, and Energy consumption of user device}
We have considered different tasks in our experiment. The FL-based model is used for offloading decision-making. At first the model is used to check whether the requested task is computationally intensive or not based on the input features such as task type and the input. If the task is predicted as computationally intensive, then, based on the input features such as uplink traffic, downlink traffic, bandwidth, latency, etc., the decision is taken whether to offload or locally execute it. For the tasks with simple computation usually local execution is performed. However, sometimes the user may need to remotely access the content or result of computation. Thus, we have considered user preference also when a request is received. If the user looks for remote access, the task is offloaded to the cloud through the edge server. Otherwise, based on the outcome of the FL-based decision-making model, the task is either locally executed or offloaded to the edge server. The response time for the considered task execution scenarios and the energy consumption of the user device during the period, are presented in Table \ref{tab:offresult}. 
\par
As the \textit{first task type}, we have considered a basic calculator design that performs four operations: addition, subtraction, multiplication, and division. Most of the user devices contain this basic application, which can work irrespective of the Internet connectivity. The user inputs numbers, and the calculator generates the results according to the requested operation. As the \textit{second task type}, we have considered matrix multiplication, where based on the order taken as input from the user, it is decided whether it would be computationally intensive or not. If the prediction result is \textit{computationally intensive}, the FL-based decision-making model is used to predict whether to offload or not based on the present network throughput, uplink and downlink traffic, etc. Based on the prediction result, the task is either offloaded or locally executed. Here, we have considered four different cases. For the first case, the two matrices are of order 50x50, and based on the prediction result, the task has been executed locally. The response time and energy consumption of the device during the period are measured, and presented in Table \ref{tab:offresult}. For the other three cases, based on the prediction results, the tasks are offloaded. The response time and energy consumption of the device during the period are measured, and presented in Table \ref{tab:offresult}. As the \textit{third task type}, we have considered file creation. The file has been created locally or remotely, according to the user preference. We have measured the response time and energy consumption of the device during the period, and presented the results in Table \ref{tab:offresult}. 
\begin{table*} 
    \centering
    \caption{Response time and energy consumption of the user device in FLDec-based task execution}
    \begin{tabular}{|c| c| c| c| c|}
\hline
 \multicolumn{2}{|c|}{Task} & Decision & Response Time (s) & \multicolumn{1}{c|}{Energy consumption (J)}\\
 \multicolumn{2}{|c|}{}& (Offload/Local)& & \multicolumn{1}{c|}{of user device}\\
\hline
\multirow{4}{*}{(1) Calculator} & 
   \multicolumn{1}{c|}{Addition: Five digit numbers}& \multicolumn{1}{c|}{Local}
   & \multicolumn{1}{c|}{0.01 (Offload: 0.1)}  & \multicolumn{1}{c|}{0.05 (Offload: 0.5)}\\
   \cline{2-5}
 & 
   \multicolumn{1}{c|}{Subtraction: Five digit numbers}& \multicolumn{1}{c|}{Local}
   & \multicolumn{1}{c|}{0.0065 (Offload: 0.125)} & \multicolumn{1}{c|}{0.0325 (Offload: 0.625)}\\
   \cline{2-5}
   & 
   \multicolumn{1}{c|}{Multiplication: Five digit numbers}& \multicolumn{1}{c|}{Local}
   & \multicolumn{1}{c|}{0.018 (Offload: 0.128)} & \multicolumn{1}{c|}{0.09 (Offload: 0.64)}\\
    \cline{2-5}
   & 
   \multicolumn{1}{c|}{Division: Five digit numbers}& \multicolumn{1}{c|}{Local}
   & \multicolumn{1}{c|}{0.005 (Offload: 0.127)} & \multicolumn{1}{c|}{0.025 (Offload: 0.635)}\\
\hline
\multirow{4}{*}{(2) Matrix multiplication} & 
\multicolumn{1}{c|}{Order: 50x50, 50x50}& \multicolumn{1}{c|}{Local}
   & \multicolumn{1}{c|}{0.06 (Offload: 0.14)}  & \multicolumn{1}{c|}{0.3 (Offload: 0.7)}\\
   \cline{2-5}
&
   \multicolumn{1}{c|}{Order: 100x100, 100x100}& \multicolumn{1}{c|}{Offload}
   & \multicolumn{1}{c|}{0.46 (Local: 0.6684)}  & \multicolumn{1}{c|}{2.3 (Local: 3.342)}\\
   \cline{2-5}
 & 
   \multicolumn{1}{c|}{Order: 200x200, 200x200}& \multicolumn{1}{c|}{Offload}
   & \multicolumn{1}{c|}{4.55 (Local: 5.78)} & \multicolumn{1}{c|}{22.75 (Local: 28.9)}\\
   \cline{2-5}
   & 
   \multicolumn{1}{c|}{Order: 300x300, 300x300}& \multicolumn{1}{c|}{Offload}
   & \multicolumn{1}{c|}{23.476 (Local: 26.43)} & \multicolumn{1}{c|}{117.38 (Local: 132.15)}\\
\hline
\multirow{6}{*}{(3) File creation} & 
   \multicolumn{1}{c|}{User preference: Local, size: 5KB}& \multicolumn{1}{c|}{Local}
   & \multicolumn{1}{c|}{1.92}  & \multicolumn{1}{c|}{9.6}\\
   \cline{2-5}
 & 
   \multicolumn{1}{c|}{User preference: Remote, size: 5KB}& \multicolumn{1}{c|}{Offload}
   & \multicolumn{1}{c|}{1.37} & \multicolumn{1}{c|}{6.85}\\
   \cline{2-5}
& 
   \multicolumn{1}{c|}{User preference: Local, size: 9KB}& \multicolumn{1}{c|}{Local}
   & \multicolumn{1}{c|}{2.07} & \multicolumn{1}{c|}{10.35}\\
   \cline{2-5}
& 
   \multicolumn{1}{c|}{User preference: Remote, size: 9KB}& \multicolumn{1}{c|}{Offload}
   & \multicolumn{1}{c|}{1.88} & \multicolumn{1}{c|}{9.4}\\  
   \cline{2-5}
   & 
   \multicolumn{1}{c|}{User preference: Local, size: 28KB}& \multicolumn{1}{c|}{Local}
   & \multicolumn{1}{c|}{2.72} & \multicolumn{1}{c|}{13.6}\\
   \cline{2-5}
   & 
   \multicolumn{1}{c|}{User preference: Remote, size: 28KB}& \multicolumn{1}{c|}{Offload}
   & \multicolumn{1}{c|}{2.61} & \multicolumn{1}{c|}{13.05}\\  
\hline
\end{tabular}
    \label{tab:offresult}
\end{table*}
\par
For evaluating the performance for a wide range of users, we performed a simulation in MATLAB2024. The average data transmission speed considered in the simulation is 1000Mbps, and the task size is randomly chosen from the range 10,000-50,000 bits. In Fig. \ref{avgres}, the average response time for 100-1000 users' requests for task execution are presented. The average energy consumption of the user device during the periods are also computed and presented in Fig. \ref{avgener}. As we observe the average response time lies in the range of 3-4.25s, and the average energy consumption lies in the range of 14-22J. 
\par
\textit{Comparison of task offloading using FLDec with baselines:} As the baseline for comparison, we considered two scenarios: (i) all tasks are offloaded, (ii) all tasks are locally executed. We observe from Table \ref{tab:offresult} that for the task \textit{Calculator}, the proposed scheme \textit{FLDec} suggested \textit{local execution}. If the task has been offloaded, then the response time and energy consumption for addition, subtraction, multiplication, and division operations would be $\sim$90\%, $\sim$95\%, $\sim$85\%, and $\sim$96\%, higher than the local execution. For the task matrix multiplication, four scenarios are considered as presented in Table \ref{tab:offresult}. For the first scenario, the FLDec has decided \textit{local execution}, and it has $\sim$57\% lower time and energy consumption than offloading it. For the other three scenarios, FLDec has suggested \textit{offload}, and we have observed that the response time and energy consumption both are reduced by $\sim$11-31\% than local execution. Hence, we observe that if all the considered tasks are locally executed, then for higher order matrix multiplication, the response time and energy consumption would be high. Similarly, if all the considered tasks are offloaded, for calculator, the response time and energy consumption would be high for all of the considered operations, and for matrix multiplication the first scenario would take more time and the energy consumption also would be high. Thus, from the results we observe that the proposed FL-based decision-making model has taken appropriate decisions for the considered case studies, and reduced the response time and energy consumption.  
\begin{figure} 
\centering
    \includegraphics[width=0.995\linewidth, height=2.8in]{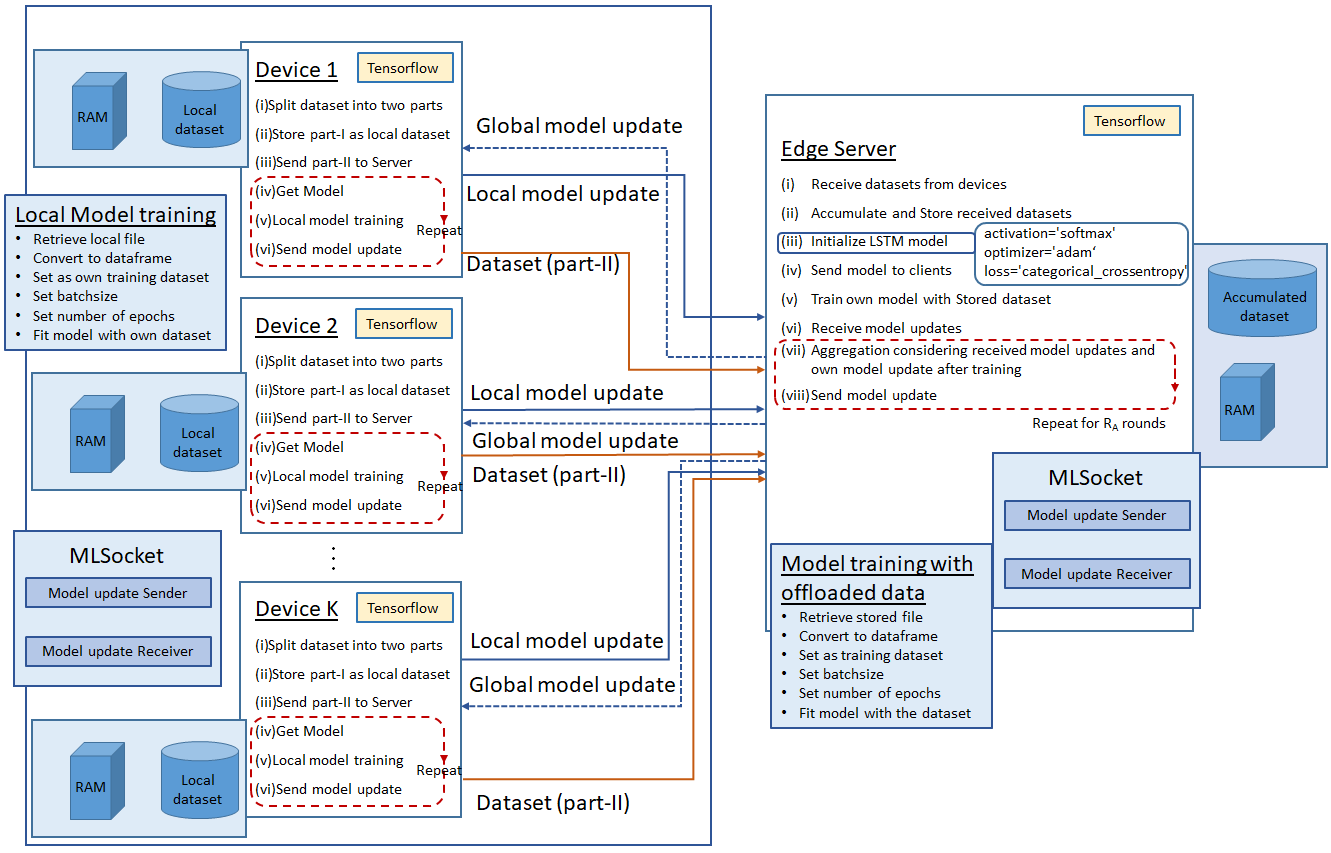}
    \caption{\centering{Implementation diagram of the proposed federated offloading framework}}
    \label{implement}
\end{figure}

\subsection{Performance of FedOff}
The performance of the proposed partial computation offloading method for FL which is referred to as federated offloading or \textit{FedOff}, is evaluated in terms of prediction accuracy, precision, recall, and F1-Score of the global as well as local models, total time consumption, and energy consumption of the user device during the period. The implementation diagram of the proposed framework is presented in Fig. \ref{implement}.

\begin{figure} 
\centering
    \includegraphics[width=0.9\linewidth, height=2.5in]{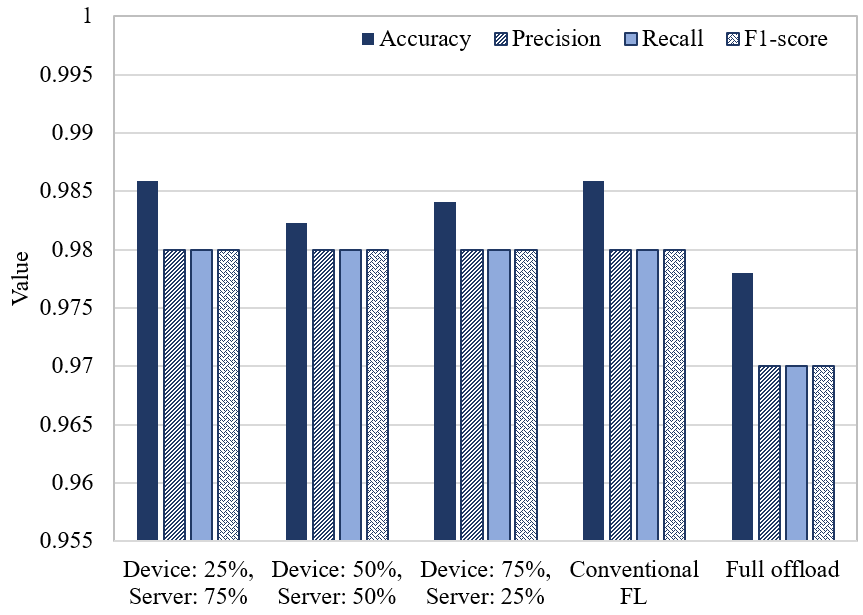}
    \caption{\centering{Accuracy, precision, recall, and F1-Score of the global model in conventional FL, federated offloading, and full offloading}}
    \label{global_fo}
\end{figure}
\begin{figure} 
\centering
    \includegraphics[width=0.9\linewidth, height=2.5in]{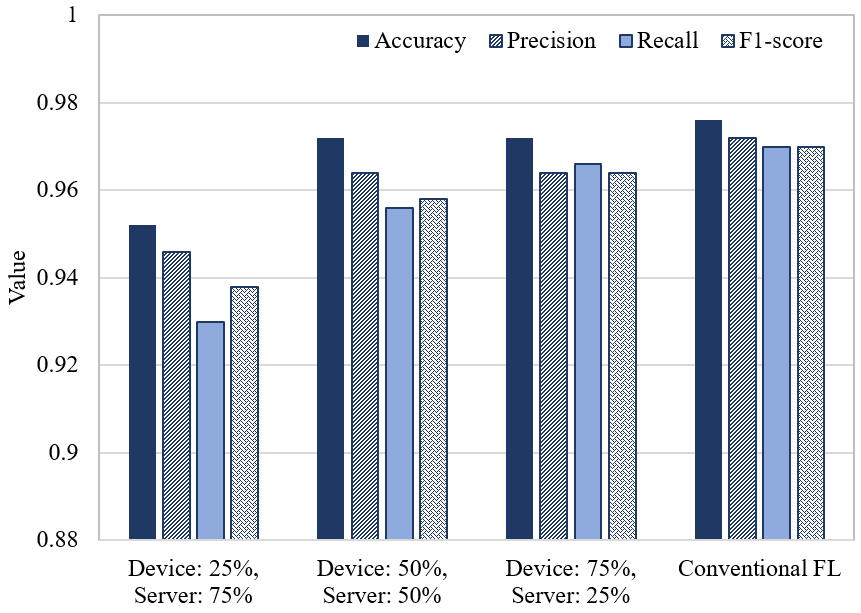}
    \caption{\centering{Average accuracy, precision, recall, and F1-Score of the local models in conventional FL and federated offloading}}
    \label{local_fo}
\end{figure}
\subsubsection{Experimental setup}
To evaluate the performance of \textit{FedOff}, we have provisioned six virtual machines from the RONIN cloud environment. Among them five act as the clients, and one acts as the edge server. Each of the clients has 4GB memory. The edge server has 4GB memory and the processor is Intel(R) Xeon(R) Platinum 8259CL CPU @ 2.50GHz. The number of rounds was set to 5, and $\alpha_r=1$. To evaluate the performance of the proposed federated offloading method, we have considered an activity recognition dataset\footnote{\url{https://www.kaggle.com/datasets/nurulaminchoudhury/harsense-datatset}}. Activity recognition is one of the popular application used by mobile users. Thus, we consider this application in our work. 
\begin{figure*}
    \centering
    \begin{subfigure}[b]{0.3\textwidth}
        \centering
        \includegraphics[width=0.99\linewidth, height=2.0in]{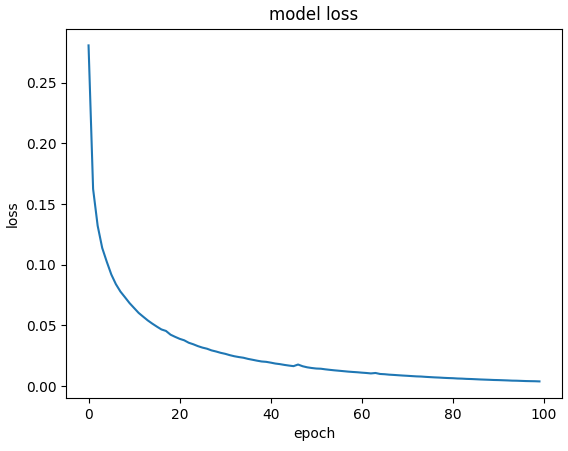}
        \caption{\centering{Global model loss for case 1 (local-25\%, offload-75\%)}}
        \label{loss1}
    \end{subfigure}
    \hfill
    \begin{subfigure}[b]{0.3\textwidth}
        \centering
        \includegraphics[width=0.99\linewidth, height=2.0in]{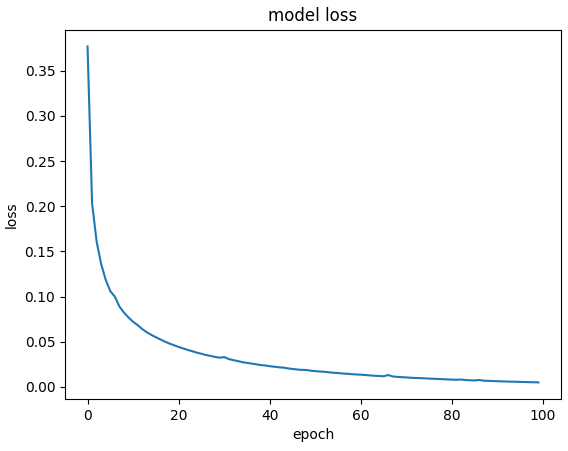}
        \caption{\centering{Global model loss for case 2 (local-50\%, offload-50\%)}}
        \label{loss2}
    \end{subfigure}
    \hfill
    \begin{subfigure}[b]{0.3\textwidth}
        \centering
        \includegraphics[width=0.99\linewidth, height=2.0in]{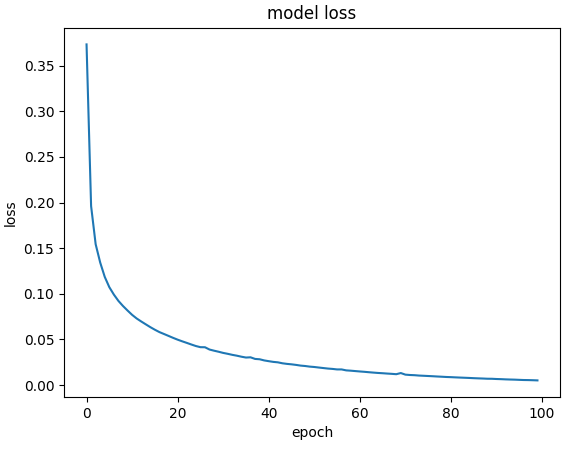}
        \caption{\centering{Global model loss for case 3 (local-75\%, offload-25\%)}}
        \label{loss3}
    \end{subfigure}
    \caption{Global model loss in proposed federated offloading method}
    \label{loss}
\end{figure*}
\subsubsection{Prediction accuracy}
The dataset has been split among five user devices as their local datasets and the server as the global dataset. The clients i.e. devices split their datasets into the following ratios: (i) local-25\%, offload-75\%, (ii) local-50\%, offload-50\%, and (iii) local-75\%, offload-25\%. The portion of data to be offloaded has been encrypted using Fernet cryptography, and then sent to the server. Fernet is a symmetric key cryptography method, which is based on AES-128 for encryption and HMAC using SHA256 for authentication. As symmetric key cryptography is used, the same key is used for encryption as well as decryption. However, all the devices have different keys. As the latency is vital, we have considered symmetric key cryptography in FedOff that can protect data security but at lower time consumption. We have measured the latency in data transmission from the experiment. 
\begin{figure*} 
    \centering
\begin{minipage}{0.495\linewidth}
\includegraphics[width=0.99\linewidth, height=2.4in]{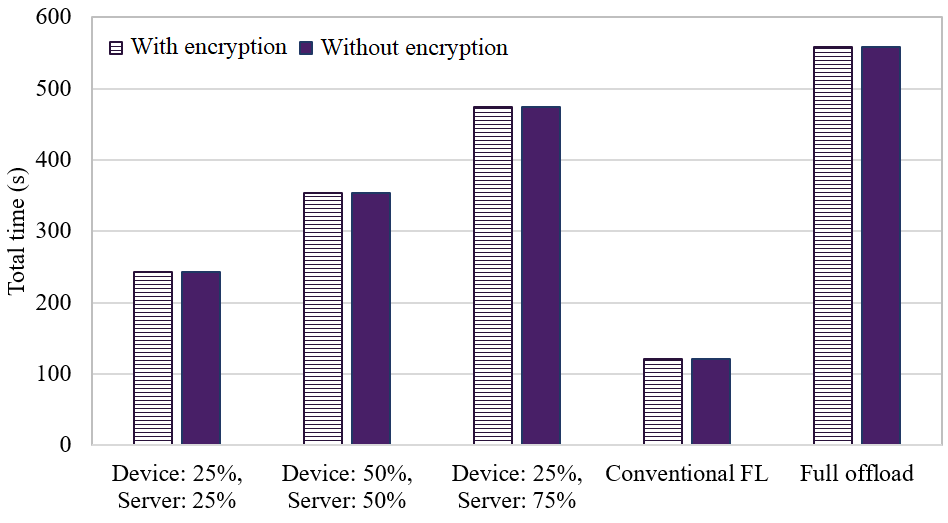}
    \caption{\centering{Total time consumption in FedOff}}
    \label{timeoff}
\end{minipage}
\begin{minipage}{0.495\linewidth}
\includegraphics[width=0.99\linewidth, height=2.4in]{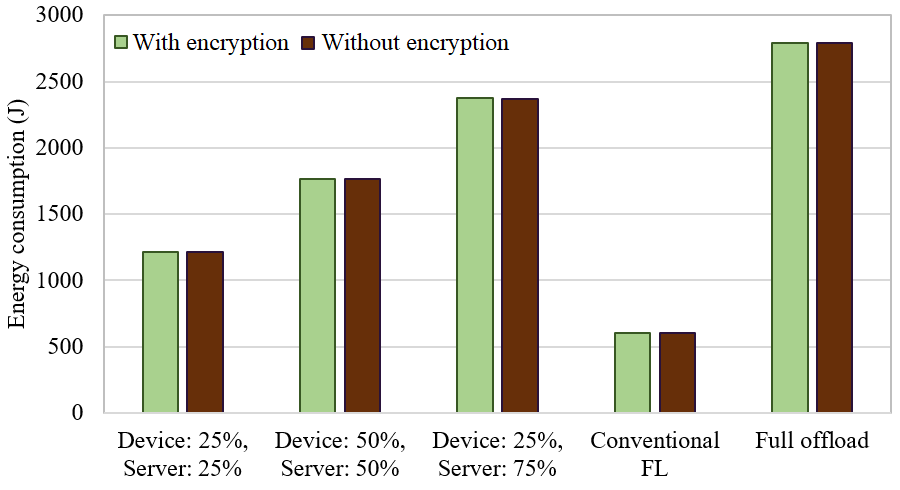}
    \caption{\centering{Average energy consumption of the participating devices during the total time in FedOff}}
    \label{energyoff}
\end{minipage}
\end{figure*}
\par
The accuracy, precision, recall, and F1-Score of the global model, achieved by FedOff are presented in Fig. \ref{global_fo}. The average accuracy, precision, recall, and F1-Score of the local models, using FedOff are presented in Fig. \ref{local_fo}. The results show that FedOff has achieved $>$98\% accuracy for the global model and $>$94\% accuracy for the local models. The precision, recall, and F1-score are 0.98 for the global model, and $>$0.92 for the local models. We have also conducted the experiment for conventional FL, where no data offloading takes place, and only model updates are exchanged between the clients and the server. We have also conducted the experiment for the case where the devices are unable to train models, and the data from the devices are fully offloaded to the edge server. The edge server trains the model for individual datasets, and then performs aggregation. After developing the model, the server sends it to the clients, so that they can perform prediction using the model in future. The accuracy, precision, recall, and F1-Score of the global model for conventional FL and full offloading are presented in Fig. \ref{global_fo}. The average accuracy, precision, recall, and F1-Score of the local models for conventional FL are presented in Fig. \ref{local_fo}. As we observe, the accuracy, precision, recall, and F1-Score for the global model in the proposed method are almost same as that of conventional FL, and better than the case of full offloading. 
\par
The global model loss for FedOff for the considered three scenarios ((i) local-25\%, offload-75\%, (ii) local-50\%, offload-50\%, and (iii) local-75\%, offload-25\%) are presented in Fig. \ref{loss}. We observe that for each of the case, the loss tends to 0, which along with the nature of the curve indicates that the model has converged. 

\subsubsection{Time and Energy consumption}
The total time for FedOff for the three considered scenarios ((i) local-25\%, offload-75\%, (ii) local-50\%, offload-50\%, and (iii) local-75\%, offload-25\%) are measured as shown in Fig. \ref{timeoff}. The total time includes the total training time (training time plus communication time for model updates exchange) considering all rounds, the time consumption for data transmission, and the time consumption for data encryption and decryption. The total time for the conventional FL and full offloading are also measured, and presented in the figure.  The average energy consumption of the devices during the total period are presented in Fig. \ref{energyoff}. As we observe the total time and energy consumption are lowest in conventional FL, where the devices locally analyse the full dataset, and exchange model updates with the server. However, if the devices cannot analyse the full dataset, then case 1 consumes the lowest time and energy consumption. Case 2 can also be adopted, where 50\% data is locally analysed, and rest 50\% is offloaded to the server. If the device is unable to analyse the dataset at all, then the entire dataset is offloaded to the server. We observe from Figs. \ref{timeoff} and \ref{energyoff} that the use of cryptography for privacy protection during data transmission consume a very less amount of time and energy. We observe that the use of encryption and decryption increases the time and average energy consumption by only 0.05-0.16\% but ensures data privacy protection during transmission. 

\subsection{Comparison with Existing Offloading Methods}
This section compares the proposed approaches with the existing offloading schemes. 
\begin{figure*} 
    \centering
\begin{minipage}{0.495\linewidth}
\includegraphics[width=0.99\linewidth, height=2.4in]{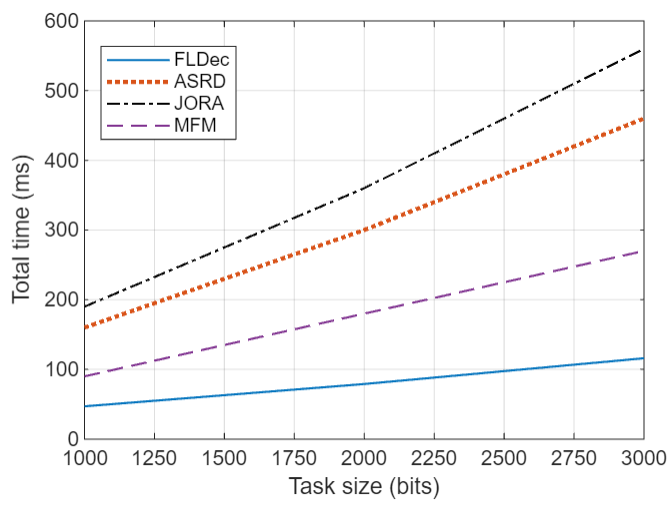}
    \caption{\centering{Total time in task offloading in FLDec and existing offloading schemes}}
    \label{comfedtime}
\end{minipage}
\begin{minipage}{0.495\linewidth}
\includegraphics[width=0.99\linewidth, height=2.4in]{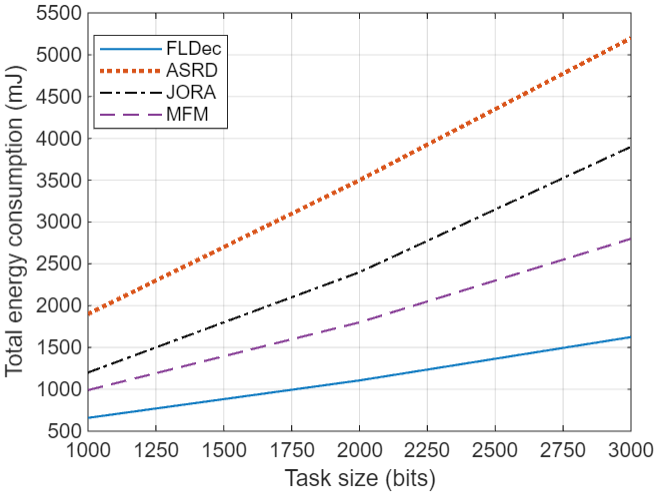}
    \caption{\centering{Total energy consumption in FLDec and existing offloading schemes}}
    \label{comfedener}
\end{minipage}
\end{figure*}
\par
\textbf{Comparison with existing offloading schemes using simulation:} In Fig. \ref{comfedtime}, the total time in task execution using FLDec is compared with the task execution time in existing offloading schemes \cite{jiang2022joint, ho2020joint, tang2024federated}, with respect to the task size. For comparison, we have performed a simulation based on different task sizes (1000-3000 bits), while performing task offloading or local execution using FLDec, and determined the time consumption in task execution. The total energy consumption (device and edge server) during task execution period is also determined and compared with the existing offloading schemes in Fig. \ref{comfedener}, with respect to the task size. As we observe from the results, FLDec has $\sim$75-79\%, $\sim$47-57\%, and $\sim$66-74\%, less time consumption than joint task offloading and resource allocation method (JORA) \cite{jiang2022joint}, multi-feature matching scheme (MFM) \cite{tang2020deep}, and action space recursive decomposition method (ASRD) \cite{ho2020joint}, respectively. We also observe that FLDec has $\sim$33-42\%, $\sim$45-58\%, and $\sim$65-68\% less energy consumption than MFM, JORA, and ASRD, respectively. In FLDec, as we have used an FL-based decision-making model to decide whether to offload or not based on task type, user input, network parameters, etc., the accurate decision-making is achieved, and the time consumption as well as energy consumption in task execution are minimal. Finally, we observe that FLDec outperforms the existing offloading approaches with respect to total time and energy consumption in task execution.
\par
\textbf{Comparison with existing schemes for offloading in FL using experiments:} We carried out a comparative study between the proposed partial computation offloading method FedOff and the existing offloading methods for FL. As the datasets used are different, we have considered the contributions for comparison along with prediction accuracy and training time. The comparative study is presented in Table \ref{tab:comres}, and we observe that the model loss was 0.2-0.5 in \cite{ji2021computation}, and the system delay was 10-50s per round. The number of rounds in \cite{ji2021computation} was 100 to converge and get minimum loss. Hence, the total delay was 1000-5000s in \cite{ji2021computation}, considering all rounds. The prediction accuracy was 80\% for \cite{wu2022fedadapt}, and the training time was 343-701s per round. The total number of rounds in \cite{wu2022fedadapt} was 100 to converge. Hence, the total training time was 34300-70100s in \cite{ji2021computation}, considering all rounds. As we observe, FedOff achieved $>$98\% accuracy for the global model and $>$94\% accuracy for the local models. The number of rounds to converge and get minimum loss in FedOff was 5. The total time (training time considering all rounds+data transmission time+time consumption for data encryption and decryption) was 200-500s in FedOff, and the average training time per round was 40-100s. The average energy consumption of the participating user devices during the total training period was 1000-2500J. Unlike our proposed method, related works \cite{wu2022fedadapt, ji2021computation} have not considered the energy consumption and secure data offloading aspects. 
\par
\begin{table*}
\caption{\centering{Comparison of FedOff with existing offloading methods for FL}}
    \centering
    \begin{tabular}{|c| c| c| c| c| c|}
        \hline
         Work & Accuracy/ Loss & Time/Delay& Energy consumption & Use of & Number of\\
         & & & of user device & cryptography & rounds to converge\\
         & & & & for data privacy & \\
         & & & & protection & \\
        \hline
  Ji et al. \cite{ji2021computation} & Loss: 0.2-0.5 & 10-50s (per round) & Not measured & \xmark & 100\\
  & & Total: 1000-5000s (100 rounds) & & &\\
   \hline
   Wu et al. \cite{wu2022fedadapt} & Accuracy: 80\% & 343-701s (per round) & Not measured & \xmark & 100\\
 & & Total: 34300-70100s (100 rounds)& & &\\
   \hline
  FedOff & Accuracy: $>$98\% (global model), & 40-100s (per round), & 1000-2500J & \cmark & 5\\
(proposed)  & $>$94\% (local models), & Total: 200-500s (training for &  & &\\
&Loss: $<$0.35  & 5 rounds, data transmission, & & &\\
& & encryption, decryption) & & &\\
       \hline
    \end{tabular}
    \label{tab:comres}
\end{table*}

\section{Conclusion and Future Work}
\label{con}
This paper proposes an FL-based offloading decision-making model for mobile devices. Based on the task type and the user input, the FL-based decision-making model predicts whether the task is computationally intensive or not, using MLP. If the predicted result is \textit{computationally intensive}, then the proposed FL-based decision-making model predicts \textit{whether to offload or locally execute} the task based on the network parameters, using LSTM. Based on the predicted result the task is either locally executed or offloaded to the edge server. If the edge server is unable to execute it or if the user preference is remote execution, then, the task is offloaded to the cloud. The experimental results present that the proposed offloading decision-making method has achieved above 90\% prediction accuracy, and the considered case scenarios show that the proposed method provides the result at lower response time and lower energy consumption of the user device. The results also present that the proposed offloading method has reduced the response time by $\sim$11-31\% for computationally intensive tasks. We have also proposed a partial computation offloading method for FL in this paper, where the user devices offload a part of their local dataset to the edge server if they are unable to analyse a huge dataset. For secure data transmission at lower time symmetric key cryptography has been used. The experimental results show that the proposed partial computation offloading method for FL has achieved $>$98\% accuracy for the global model and $>$94\% accuracy for the local models. The results also present that the use of cryptography increases the time by 0.05-0.16\% but ensures data privacy protection during transmission. 
\par
As part of our future work, we will extend the proposed work to enhance prediction accuracy in situations such as when the user's requested task does not fall in the trained task category or the input provided for the task fall apart from the training dataset. Generative artificial intelligence can be used with FL for providing a good prediction model if the training dataset has a very limited number of samples.
\bibliography{ref}
\bibliographystyle{IEEEtran}

\vskip -0.05\baselineskip plus -1fil
\begin{IEEEbiography}[\vspace{-1cm}{\includegraphics[width=1in,height=1in,clip,keepaspectratio]{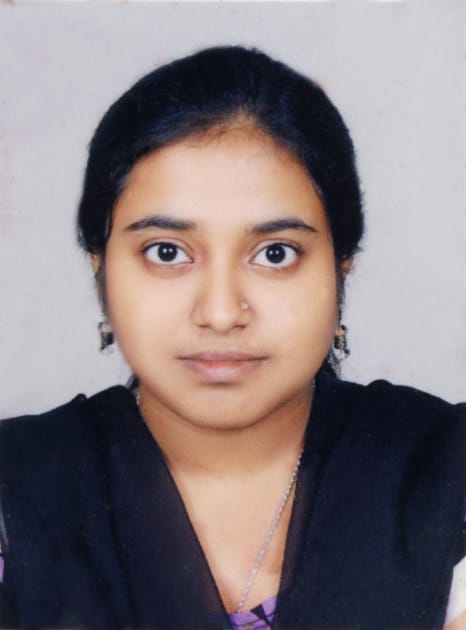}}]
    {Anwesha Mukherjee} is an Assistant Professor of the Department of Computer Science, Mahishadal Raj College, West Bengal, India. She is also a Research visitor in the CLOUDS Lab, The University of Melbourne, Australia. Her research interests include mobile computing, IoT, cloud computing, edge computing, machine learning, and federated learning.
\end{IEEEbiography}%
\vskip -1.5\baselineskip plus -1fil
\begin{IEEEbiography}[\vspace{-1cm}{\includegraphics[width=1in,height=1in,clip,keepaspectratio]{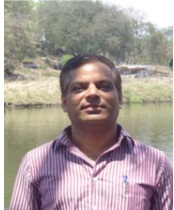}}]
    {Rajkumar Buyya} (Fellow, IEEE) is a Redmond Barry distinguished professor and director of the Cloud Computing and Distributed Systems (CLOUDS) Laboratory, The University of Melbourne, Australia. He has authored more than 800 publications and seven text books including “Mastering Cloud Computing” published by McGraw Hill, China Machine Press, and Morgan Kaufmann for Indian, Chinese and international markets respectively. He is one of the highly cited authors in computer science and software engineering worldwide (h-index=168, g-index=371, 151,200+ citations). 
\end{IEEEbiography}%
\end{document}